\documentclass[aps,physrev,preprint,superscriptaddress]{revtex4-2}

\usepackage{graphicx}

\usepackage{xcolor,color}
\usepackage[normalem]{ulem}

\begin{document}


\title{Effectiveness of Quota Policies Across STEM, Biological, and Humanities Programs}


\author{R. D. Matheus}
\affiliation{Institute for Theoretical Physics, S\~ao Paulo State University (Unesp), Rua Dr. Bento Teobaldo Ferraz, 271-Bloco II, 01140-070 S\~ao Paulo, SP, Brazil}

\author{E. M. Gennaro}
\affiliation{Department of Aeronautical Engineering, Engineering School, S\~ao Paulo State University (Unesp), Avenida Professora Isette Corr\^ea Font\~o, S\~ao Jo\~ao da Boa Vista 13876-750, SP, Brazil} 

\author{M. T. Yamashita}
\affiliation{Institute for Theoretical Physics, S\~ao Paulo State University (Unesp), Rua Dr. Bento Teobaldo Ferraz, 271-Bloco II, 01140-070 S\~ao Paulo, SP, Brazil}


\date{\today}

\begin{abstract}
We examine more than a decade of quota policy at Unesp, analyzing Physics, Biology, and Pedagogy as representative programs of distinct assessment styles. Quotas show little impact in Physics, where the admission barrier is low, and in Pedagogy, where high pass rates make it difficult to differentiate students, but they reveal systematic differences in Biology. Focusing the analysis on Calculus I -- an introductory course in Physics and other Science, Technology, Engineering, and Mathematics (STEM) programs -- for which much larger statistics are available, a clear hierarchy emerges: students admitted through open competition perform best, those from public schools achieve intermediate results, and students from racial quotas perform worst. When students are divided directly by the admittance exam grade, the performance difference is even clearer. Statistical analysis also shows that, contrary to expectation, the probability of passing decreases as the number of attempts increases, indicating that initial educational gaps are difficult to overcome within higher education.
\end{abstract}


\maketitle

\section{Introduction}

Affirmative action policies, especially quotas in higher education, have long been employed as instruments to promote social inclusion. Originating in the mid-20th century, it initially focused on racial groups and later expanding to encompass gender, socioeconomic status, and other forms of presumed disadvantage. The underlying rationale is that compensatory mechanisms should correct structural imbalances by ensuring that the composition of the student body reflects, to some extent, the demographic distribution of some elected groups in society.

Particularly in Brazil, the adoption of quota policies in public universities had its landmark in 2012, when Federal Quota Law mandated that federal institutions reserve a portion of their admissions for students from public schools, with subquotas for racial groups (Black, Brown and Indigenous people). Considering the three public state universities at S\~ao Paulo, responsible for more than 1/3 of all research produced in Brazil, the first one to adopt such a policy was S\~ao Paulo State University (Unesp) in 2013, with the first cohort of quota students entering in 2014, followed by University of Campinas (Unicamp) and University of S\~ao Paulo (USP), both in 2017.

Despite the clash between political rival groups for and against quotas, which is not related to the scope of this article, there are still few direct analyses of the social efficiency of such policies or even on the academic performance of those benefited by them, comparing quota and non-quota students across different areas of knowledge (the Humanities, the Physical and Mathematical sciences, and the Biological sciences). Some articles consider that achieving demographic proportionality of Black, Brown, and Indigenous students in universities could indicate the success of affirmative action policies~\cite{Melo2025,FrancisTan2024,Vieira2019,Machado2025}. However, this evaluation may be too narrow, as the long-term effectiveness of these policies must be assessed not only in terms of access, but also through outcomes such as academic performance and graduation rates, which necessarily precede successfully entering into the job market or continuing academic endeavors.

Affirmative action policies have been implemented in several countries, such as the United States, India, and South Africa, each with specific historical and social contexts. In Brazil, however, quota policies were introduced much later, and in a scenario where public universities are both highly selective and central to scientific production. Within this framework, Unesp provides a particularly relevant case study: with a wide geographical distribution across S\~ao Paulo State, it offers a good opportunity to investigate the long-term academic outcomes of quota and non-quota students under comparable conditions.

Unesp is a public university with 34 units in 24 cities, including 22 in the interior of the state, one in the capital, and one on the coast. In this article, we analyze more than a decade of affirmative action policy at Unesp, comparing the performance of quota and non-quota students from 2013 to 2025. The multicampus structure, with students originating from different localities in Brazil and adapting to various regions in São Paulo, provides variability beneficial to this study, avoiding results that would be otherwise sensitive to localized effects.

A crucial aspect of evaluating academic performance is the distinction between the Humanities, the Physical and Mathematical sciences, and the Biological sciences~\cite{Garcia2015,Peixoto2016}, as Humanities tend to have high pass rates, which can mask important measurable effects of the policy. This distinction is often overlooked in many studies, which either do not separate fields of knowledge or adopt imprecise definitions of the student population under analysis, failing to establish a univocal correspondence between their classification criteria and the official definition of quota students~\cite{Galhardo2020,Salmi2020,Machado2025,Mello2022,Santos2023,Oliveira2024,Zeidan2024}.

The goal of this article is not to take a position on whether quotas should be adopted in universities, but to examine, with methodological rigor, the conclusions that can be drawn from ten years of Unesp’s quota policy, comparing the success rate of students admitted through the universal system (SU -- Sistema Universal), public school track (EP -- Escola P\'ublica), and Black, Brown, and Indigenous quota (PPI -- Pretos, Pardos e Ind\'\i genas), in early- and mid-program subjects in physics, biology, and pedagogy. We also compare the successful graduation rates in biology, pedagogy, and physics. For clarity and consistency with the data presented, we chose to keep the acronyms (SU, EP, PPI) in accordance with their original description in Portuguese.

The paper is organized as follows. The next section, \ref{methods}, details how Unesp’s quota policy is applied and characterizes each target group -- SU, EP and PPI --, explains how we obtained the data from the university’s database and sets out the assumptions underlying the analysis. Section \ref{results} presents the results. The final section, \ref{conclusion}, offers discussion and conclusions.

\section{Methods}
\label{methods}

In the admission process of Unesp, candidates may apply either through the Universal System or through the system of reserved vacancies for students from public schools. The SU corresponds to open competition, with all applicants automatically considered under this modality. In this case, the evaluation process does not identify or take into account candidates’ race, social background, ancestry, or any similar personal attribute. The reserved system is intended for students who have completed their entire high school education in Brazilian public schools. This criterion reflects the fact that, on average, public schools in Brazil provide weaker academic preparation than private schools, creating disadvantages for their graduates when competing for admission to public universities. 

Within the reserved system, there is a further subdivision for applicants who self-declare as Black, Brown, or Indigenous. Self-declared Black or Brown candidates are subject to a verification procedure focused on their phenotypical characteristics. For the purposes of this procedure, the following are considered phenotypical traits typically associated with Black or Brown individuals: dark skin color (brown or black), curly or frizzy hair texture, broad nose, and thick, brownish lips. These evaluations are carried out by heteroidentification committees, which are responsible for confirming or rejecting the declared racial/ethnic status. Only candidates whose declarations are validated remain eligible for PPI-reserved seats. For Indigenous candidates, verification is based on documentation provided by legally recognized Indigenous communities and by the federal agency responsible for Indigenous affairs, FUNAI (Funda\c c\~ao Nacional dos Povos Ind\'\i genas -- National Foundation of Indigenous Peoples).

From the total number of places offered in the entrance examination, 50\% are reserved for candidates from public schools. Within this fraction, 35\% of the total seats are specifically allocated to PPI candidates. The remaining 50\% of the seats are filled through the SU. If an EP or PPI candidate obtains a score high enough to qualify for admission through the SU, they are admitted under the SU category, and the corresponding reserved seat is then released for another EP or PPI candidate. The same logic applies to PPI candidates who achieve a score sufficient to enter through the EP category (but not high enough for SU): in such cases, they are admitted as EP, and one place is released in the PPI category. This redistribution implies, by construction, that the PPI subgroup consistently remains associated with the lowest admission cutoff scores in the examination. It also implies that the best performing students in the admission exam will be in the SU group, including many that identify in racial groups and from public schooling. The results presented below should be interpreted in that light, avoiding undue identification of the quota categories here solely with race or school system.

The overall structure of the admission system is summarized in Table~\ref{table1}.  

\begin{table}[h]
\centering
\caption{Admission systems in the Unesp entrance examination}
\label{table1}
\begin{tabular}{|l|l|}
\hline
\textbf{System} & \textbf{Description} \\ \hline
SU & Universal system, open competition. \\ \hline
EP & Public school students. \\ \hline
PPI & Public school students who self-declare as Black, Brown, or Indigenous, \\
    & subject to heteroidentification. \\ \hline
\end{tabular}
\end{table}

The data analyzed were provided by the Data Management Office at Unesp, with no personal identification of students. We considered only students admitted through the entrance exam: transfer, visiting, and other atypical cases were excluded. Final status was restricted to students who either obtained the degree or abandoned the program, with transfers excluded as well. As far as specific courses are concerned, students who withdrew mid-semester, failed due to absence, or completed the course with a grade below the minimum requirement were all classified as failures. Rare non-passing cases (e.g., documented legal or medical reasons) were excluded.

The data set, shown in Table \ref{tabledataset1}, was updated in July 2025 and includes only admissions for that year, since data from the first semester of 2025 was incomplete. Note that the sum of graduates and withdrawals differs from the number of incoming students since some students remain in an undefined status and may still either graduate or withdraw in the future.

\begin{table}[h]
\centering
\caption{Programs data set: total number of students analyzed from 2013 to July 2025.}
\label{tabledataset1}
\begin{tabular}{|l|c|c|c|}
\hline
\textbf{Program} & \textbf{Incoming Students} & \textbf{Graduates} & \textbf{Withdrawals} \\ \hline
Physics & 3188 & 341 & 2214 \\ \hline
Biology & 6946 & 1967 & 2581 \\ \hline
Pedagogy & 4409 & 1884 & 1295 \\ \hline
{\it Total} & 14543 & 4192 & 6090 \\ \hline
\end{tabular}
\end{table}

The programs analyzed in this study are offered in different cities within the Unesp multicampus system: Physics is located in six units (Bauru, Guaratinguet\'a, Ilha Solteira, Presidente Prudente, Rio Claro, and S\~ao Jos\'e do Rio Preto); Biological Sciences in eight (Assis, Bauru, Botucatu, Ilha Solteira, Jaboticabal, Rio Claro, S\~ao Jos\'e do Rio Preto, and S\~ao Vicente); and Pedagogy in six (Araraquara, Bauru, Mar\'\i lia, Presidente Prudente, Rio Claro, and S\~ao Jos\'e do Rio Preto).

The data presented in Table~\ref{calcI_dataset} were not planned a priori, but rather emerged as a natural consequence of the separate analysis of the programs, as will be explained in the next section. For this ``Calculus'' data set, we considered only students admitted through Unesp’s main entrance examination, which assigns scores from 0 to 100 and is subsequently subjected to the quota system -- all other admission types were excluded (e.g., from science olympiads). The analysis was further restricted to the period from 2015 to 2024, since scores were not consistently available for 2013 and 2014, and results from the first semester of 2025 had not yet been fully entered in the database by July 2025, when the data were collected. The criteria for inclusion in the analysis, as well as for classification into passing or failing Calculus I, were the same as those adopted for the courses in the ``Programs'' data set. Under these restrictions, information was available for 25899 students, who enrolled in Calculus I a total of 33262 times, with the results shown in Table~\ref{calcI_dataset}.

\begin{table}[h]
\centering
\caption{Calculus data set (performance in Calculus I).}
\label{calcI_dataset}
\begin{tabular}{|c|c|c|}
\hline
{Incoming students} & {Passes} & {Fails} \\ \hline
33262 & 15417 & 17845 \\ \hline
\end{tabular}
\end{table}

\section{Results}
\label{results}

In order to properly analyze the results, it is essential to evaluate the uncertainties and to distinguish among the Humanities, the Physical and Mathematical sciences, and the Biological sciences. Some studies consider only the standard deviations arising from statistical distributions~\cite{Garcia2015,Galhardo2020}, which is insufficient for a correct assessment of the problem as a whole. Other studies do not even perform a careful analysis of the uncertainties~\cite{Salmi2020,Oliveira2024}. Beyond the statistical distribution itself, it is also necessary to account for uncertainties stemming from the sample of students under evaluation. All these factors compromise a reliable assessment of the problem, since uncertainty increases as the number of candidates decreases -- as is the case when analyzing smaller classes -- and it is crucial to determine whether the groups are statistically different. Furthermore, failing to distinguish among the different areas means that the high approval rates in the Humanities, where passing or not in courses might not even be how academics differentiate good or bad students, may wash out effects that are visible in the physical, mathematical and Biological Sciences, where objetive testing is much more frequently the norm.

We assumed that the number of students ($N$) follows a Poisson distribution, with the corresponding standard deviation ($\sigma$) given by $\sqrt{N}$. In addition, a systematic uncertainty of $3\%$ from the database \cite{Barchard2020} was included in the total error bars, which were calculated using the standard error propagation method. Our analysis of the Unesp data showed that database errors, such as undue duplication or misnaming of courses, were below the 3\% error, so this value can be regarded as an upper limit.

\begin{figure}
\includegraphics[width=0.8\textwidth]{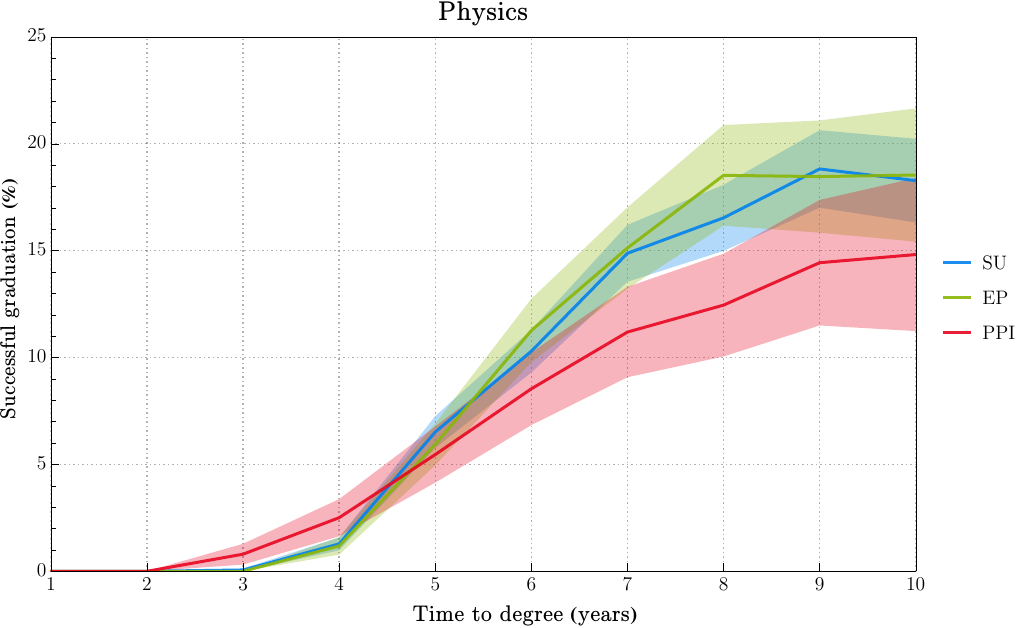}%
\caption{Average cohort evolution for the three admission tracks (defined in the main text) in the Physics programs of Unesp. The average was taken over the cohorts of year 2013 through 2023, and the curves show the average percentage of students successfully graduating after a number of years in the course. The shaded regions show the uncertainty of each curve.\label{evol_phys}}
\end{figure}

Figure \ref{evol_phys}  shows that, for the Physics programs, the three tracks are statistically indistinguishable over the entire time window: the one-$\sigma$ bands largely overlap at practically all times, and any apparent ordering among curves is not resolved by the quoted errors. The broader band for larger times reflects smaller $N$'s, since fewer cohorts have reached that number of years. Across tracks, most completions occur within roughly eight years—near the nominal time to degree—after which the curves plateau saturate at about 18$\%$ for SU and EP, and 15$\%$ for PPI. Within the resolution of this dataset, we therefore find no statistically significant track-dependent differences in cumulative graduation rates.

\begin{figure}
\includegraphics[width=0.8\textwidth]{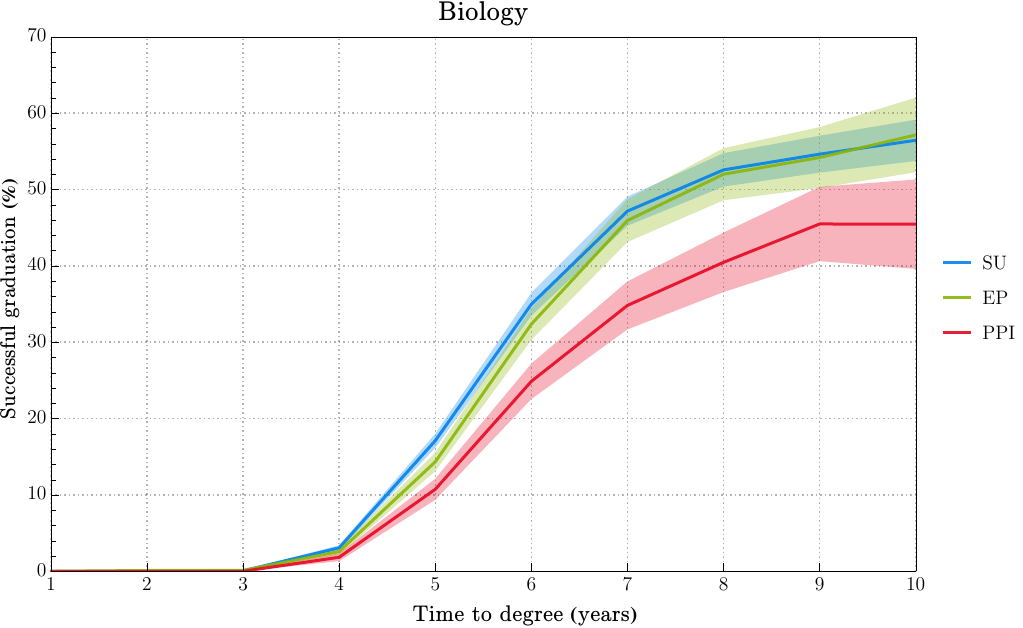}%
\caption{Average cohort evolution for the three admission tracks (defined in the main text) in the Biology programs of Unesp. The average was taken over the cohorts of year 2013 through 2023, and the curves show the average percentage of students successfully graduating after a number of years in the course. The shaded regions show the uncertainty of each curve.\label{evol_bio}}
\end{figure}

Figure \ref{evol_bio}, analogous to Fig.~\ref{evol_phys} for Physics, shows the cohort evolution in Biology. In this case, there are statistically significant differences. The one-$\sigma$ bands do not overlap over most of the time window, showing that SU and EP students reach higher graduation rates than PPI, which consistently remains at the lowest level \footnote{We stress that quota categories cannot be identified solely with race or school system. The results presented here should not be used to evaluate the academic performance of racial categories or of students from public schools.}. The curves saturate at approximately 60$\%$ for SU and EP, and 45$\%$ for PPI. These differences are robust within the quoted uncertainties, indicating track-dependent effects in Biology that were not observed in Physics.

\begin{figure}
\includegraphics[width=0.8\textwidth]{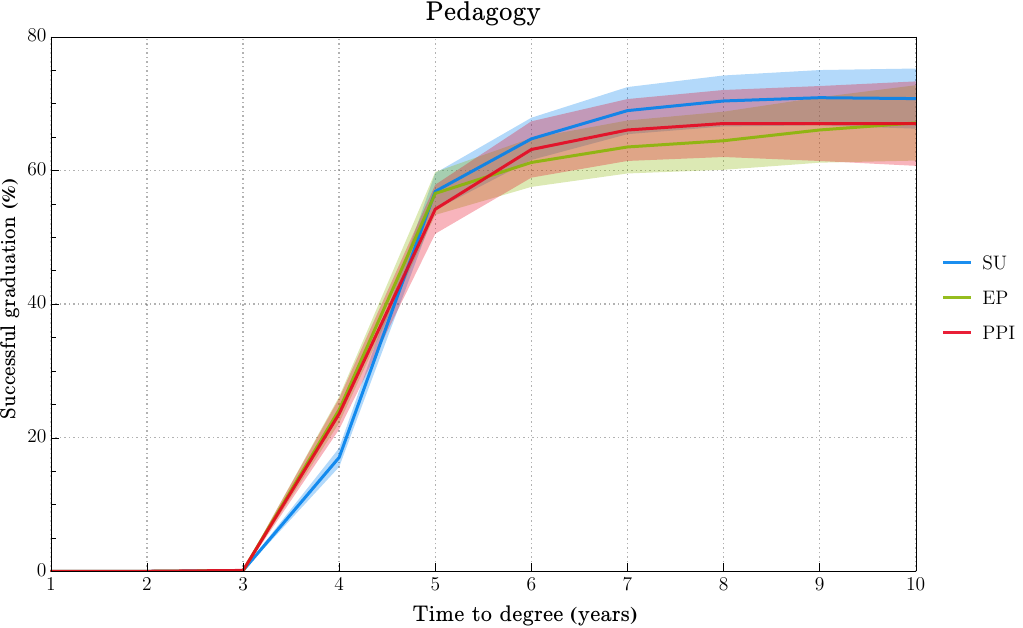}%
\caption{Average cohort evolution for the three admission tracks (defined in the main text) in the Pedagogy courses of Unesp. The average was taken over the cohorts of year 2013 through 2023, and the curves show the average percentage of students successfully graduating after a number of years in the course. The shaded regions show the statistical uncertainty of each curve.\label{evol_pedag}}
\end{figure}

Figure \ref{evol_pedag}, analogous to Figs.~\ref{evol_phys} and \ref{evol_bio}, shows the cohort evolution in Pedagogy. In this case, the plateau is much more clearly defined, and the overall success rates are considerably higher, reflecting the much lower failure rates compared to Physics and Biology -- the curves saturate after only 5 years, around the high rate of 70\%. The differences among tracks are not statistically significant.

The difference among the three programs correlates not only with the difficulty of graduating in programs -- more pronounced in Physics, intermediate in Biology, and much lower in Pedagogy -- but also with the ratio of applicants per available seat in the entrance examinations. The average applicant-to-seat ratios for the 2025 entrance exam, considering all campuses and the three modalities (full-time, daytime and evening), are $1.6 \pm 0.3$ for Physics, $4.5 \pm 0.6$ for Biology, and $2.21 \pm 0.24$ for Pedagogy. In Brazil, these examinations usually involve two stages, and the values quoted here refer to the first stage; the corresponding numbers for the second stage are typically smaller.

The very low number of applicants to Physics implies that there is no effective distinction among admission tracks. In this case, the presence or absence of quotas is essentially irrelevant, since virtually all candidates are admitted -- in some campuses the ratio of applicants per seat is even below unity. For Biology, whose demand is roughly three times larger, a statistical difference becomes visible: SU and EP students achieve higher graduation rates compared to PPI, which remains lower throughout almost all years considered. In Pedagogy, any possible differences would be washed out by the overall high success rate: the plateau is reached rapidly, with graduation probabilities approaching 70\% across all tracks.

\subsection{Analysis by courses}

In addition to analyzing the time to graduation in each program, it is also of interest to investigate whether differences arise in the success rates of specific courses within each program. To this end, we examined the performance of students admitted through SU, EP, and PPI in introductory courses of physics (Calculus I), biology (Cell Biology), and pedagogy (Philosophy of Education I), as well as in a mid-program physics course (Electromagnetism I).

\begin{figure}
\includegraphics[width=0.8\textwidth]{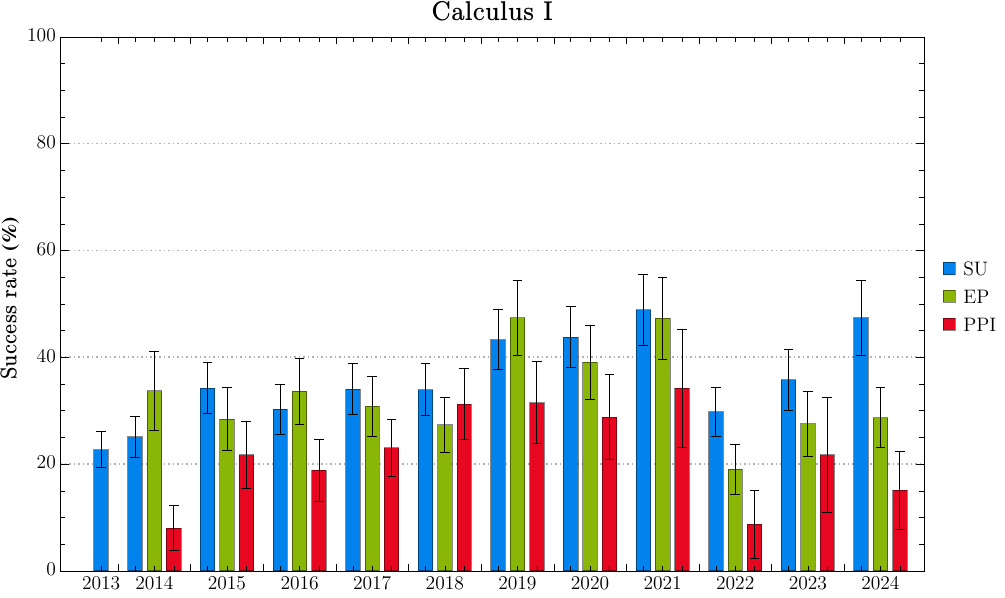}%
\caption{Ratio $P \over E$ for Calculus I in Physics courses, where $P$ is the number of students passed and $E$ is the number of students enrolled (each year and in each category). The categories represent different admission tracks, as defined in the main text.\label{aprov_calc}}
\end{figure}

Figure \ref{aprov_calc} shows the success rate in Calculus I for SU, EP, and PPI students from 2013 to 2024. The error bars correspond to one-$\sigma$ statistical uncertainties, calculated considering the deviations from a Poisson distribution and assuming a systematic error of 3\% \cite{Barchard2020}. The central values for PPI are systematically lower than those for SU and EP; however, within the quoted uncertainties, it is not possible to claim a statistically significant difference. These results, nevertheless, suggest that with larger statistics such differences might become more clearly resolved. SU and EP, on the other hand, remain statistically indistinguishable throughout the entire period.

\begin{figure}
\includegraphics[width=0.8\textwidth]{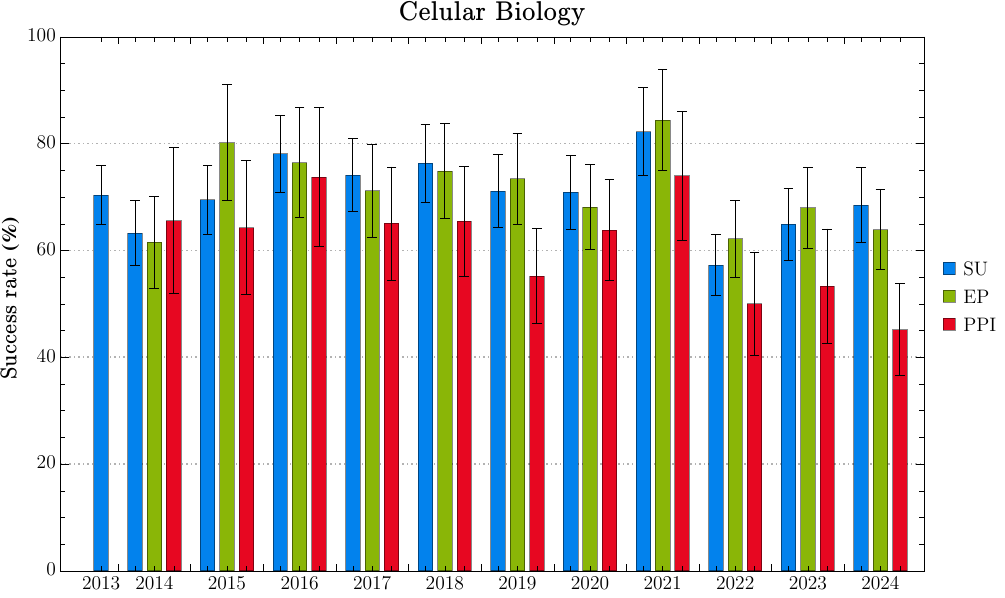}%
\caption{Ratio $P \over E$ for Celular Biology in Biology courses, where $P$ is the number of students passed and $E$ is the number of students enrolled (each year and in each category). The categories represent different admission tracks, as defined in the main text.\label{aprov_celbio}}
\end{figure}

Figure \ref{aprov_celbio}, same as Fig.~\ref{aprov_calc} but for Celular Biology, shows that the separation of PPI from SU and EP is less evident than in Calculus I. Although the central values for PPI are often lower, the uncertainties largely overlap with those of the other tracks. However, as in the previous case, these results suggest that with larger statistics potential differences might become more clearly resolved. SU and EP remain statistically indistinguishable throughout the entire period.

\begin{figure}
\includegraphics[width=0.8\textwidth]{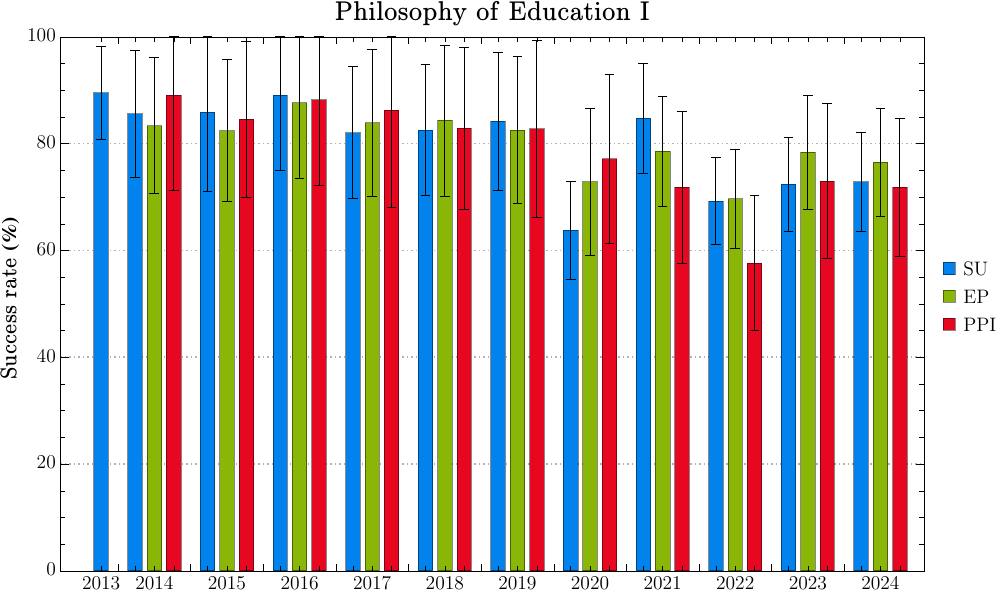}%
\caption{Ratio $P \over E$ for Philosophy of Education I in Pedagogy courses, where $P$ is the number of students passed and $E$ is the number of students enrolled (each year and in each category). The categories represent different admission tracks, as defined in the main text.\label{aprov_phyleduc}}
\end{figure}

Figure \ref{aprov_phyleduc}, same as Figs.~\ref{aprov_calc} and \ref{aprov_celbio} but for Pedagogy, shows that the error bars do not allow the identification of any performance differences among SU, EP, and PPI students. This behavior is consistent with the well-defined plateau observed in Fig.~\ref{evol_pedag}, which reflects the overall high success rate of students in the Pedagogy program.

\begin{figure}
\includegraphics[width=0.8\textwidth]{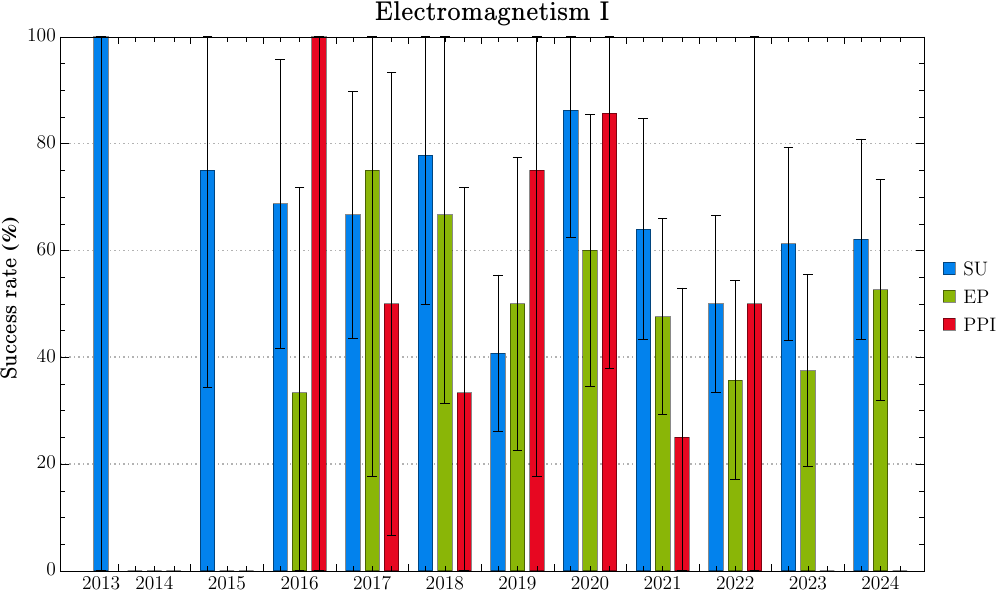}%
\caption{Ratio $P \over E$ for Electromagnetism I in Physics courses, where $P$ is the number of students passed and $E$ is the number of students enrolled (each year and in each category). The categories represent different admission tracks, as defined in the main text.\label{aprov_electroI}}
\end{figure}

Figure \ref{aprov_electroI}, same as the previous bar plots but for Electromagnetism I in the Physics program, shows that the small number of students enrolled in this advanced course leads to large statistical uncertainties. As a result, no conclusions can be drawn regarding differences among SU, EP, and PPI. Similar behavior is observed for intermediate courses in Biology and Pedagogy, where the limited statistics prevents any reliable comparison. This highlights the importance of having a sufficiently large dataset in order to assess student performance across admission tracks. 

As illustrated in Figs. \ref{aprov_calc}, \ref{aprov_celbio} and \ref{aprov_phyleduc} it is therefore more appropriate to collect data from disciplines that effectively discriminate students through failure rates. For this reason, we gathered data from all Unesp programs that include Calculus I in their curricula and recomputed the approval histogram.

\begin{figure}
\includegraphics[width=0.8\textwidth]{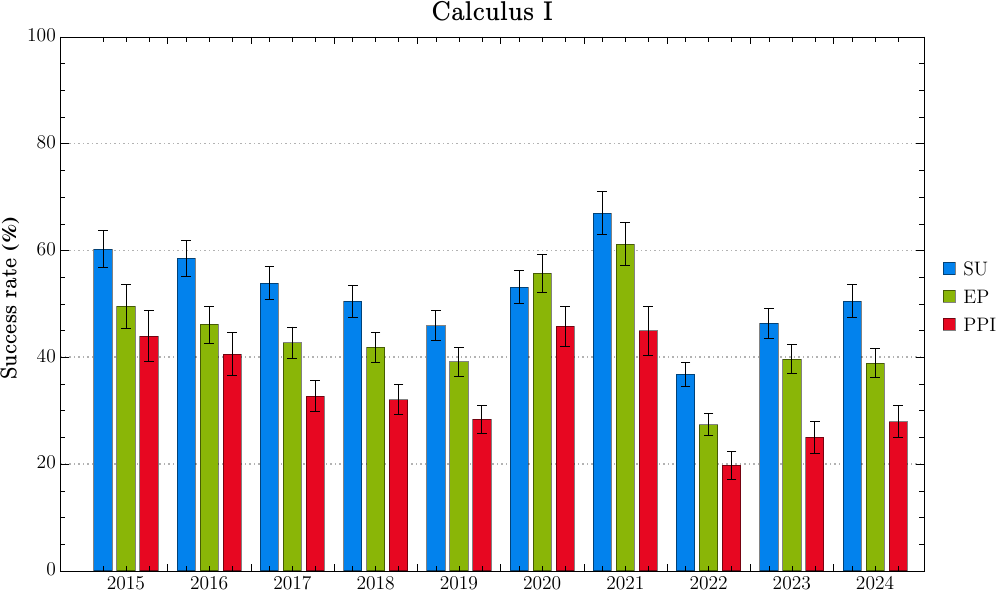}%
\caption{Ratio $P \over E$ for Calculus I in all Unesp courses, where $P$ is the number of students passed and $E$ is the number of students enrolled (each year and in each category). The categories represent different admission tracks, as defined in the main text.\label{aprov_calc_s2}}
\end{figure}

Figure \ref{aprov_calc_s2} shows the success rate in Calculus I when data from all Unesp programs offering this discipline are combined, thus providing much better statistics. In this case, a systematic difference becomes evident: SU students consistently achieve the highest success rates, EP students show intermediate performance, and PPI students remain at the lowest level. This result indicates a strong correlation between performance in the program and performance in the entrance examination. Moreover, since Calculus I is offered in most STEM programs of Unesp, the overall approval rates increase significantly when including students from programs such as Engineering, where the applicant-to-seat ratio is higher than that of Physics and, consequently, the cutoff score in the entrance examination is also higher. It is also worth noting that the years 2020 and 2021, corresponding to the pandemic period, were atypical and those years should be interpreted with caution, most exams where being held online in that period.

\begin{figure}
\includegraphics[width=0.8\textwidth]{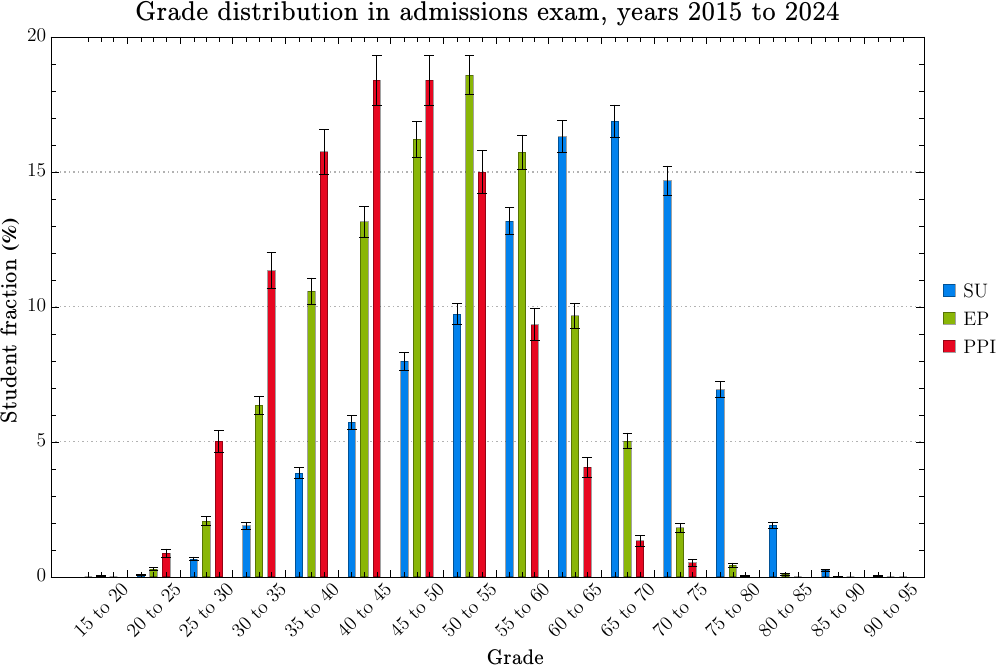}%
\caption{Grade distribution for the admission exam of all students of Unesp that took the Calculus I course between 2015 and 2024. The categories represent different admission tracks, as defined in the main text.\label{distrib}}
\end{figure}

Since admission scores are directly relevant to student performance, it is important to examine how different the grade distributions are for SU, EP, and PPI. Figure \ref{distrib} shows the distributions of admission exam grades for students admitted between 2015 and 2024. The three groups exhibit distinct profiles, consistent with the quota policy. As expected, PPI students concentrate at the lower end of the grade distribution, while SU students populate the higher ranges, with EP occupying an intermediate position. The mean and the median of the distributions are:

\vspace{0.5cm}
\noindent
Mean
\begin{itemize}
    \item SU: 60.54(10)
    \item EP: 49.91(12)
    \item PPI: 44.78(16)
\end{itemize}

\noindent
Median
\begin{itemize}
    \item SU: 62(8)
    \item EP: 50(7)
    \item PPI: 44(7)
\end{itemize}
\vspace{0.5cm}

\begin{figure}
\includegraphics[width=0.8\textwidth]{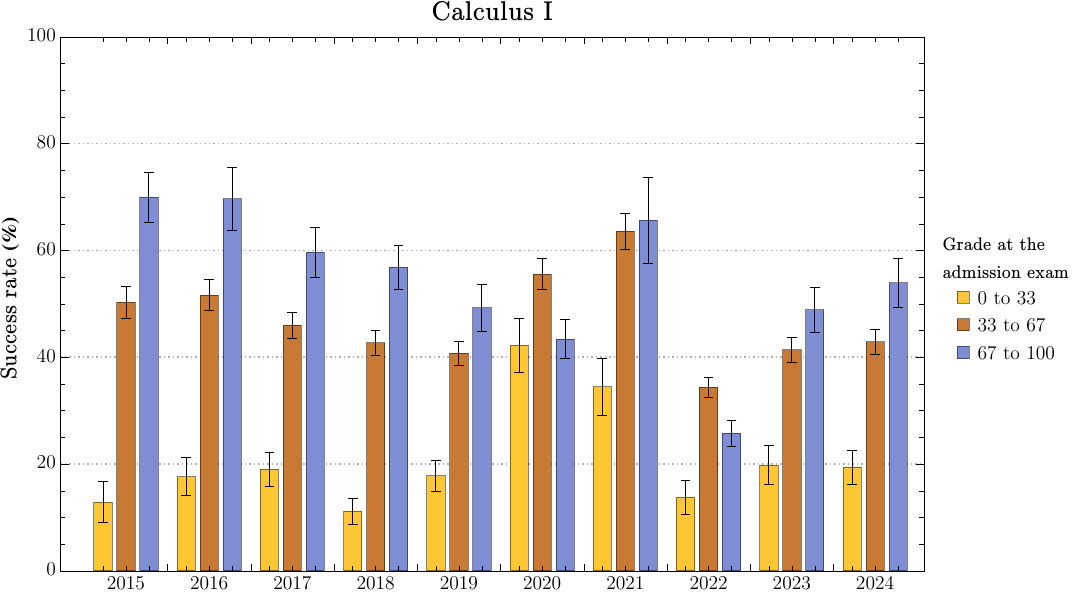}%
\caption{Ratio $P \over E$ for Calculus I in all Unesp courses, where $P$ is the number of students passed and $E$ is the number of students enrolled (each year and in each category). The categories represent different performances at the admission exams, organized by grade.\label{aprov_calc_v_grade}}
\end{figure}

Figure \ref{aprov_calc_v_grade} shows the success rate in Calculus I when students are grouped according to their admission exam scores, separated into lower, middle, and upper terciles. In this representation, the correspondence between entry performance and success in the course becomes more evident than in the division by admission track (SU, EP, PPI). The explicit grouping by admission score highlights a systematic trend (except for the pandemic period): students in the highest tercile achieve the best success rates, those in the middle tercile perform at intermediate levels, and students in the lowest tercile consistently show the lowest success. This effect naturally agrees with the differences seen in Figure \ref{aprov_calc_s2}, because of the grade distribution in Figure \ref{distrib}, yielding a consistent picture. 

The correlations in Figures \ref{aprov_calc_v_grade} and \ref{distrib} may appear at first glance to be obvious, since students with lower entrance grades also tend to perform worse in the early-year courses. One might then expect performance to improve as students progress through their programs. To test this hypothesis, we constructed a figure relating the student success rate to the number of attempts in Calculus I.

\begin{figure}
\includegraphics[width=0.8\textwidth]{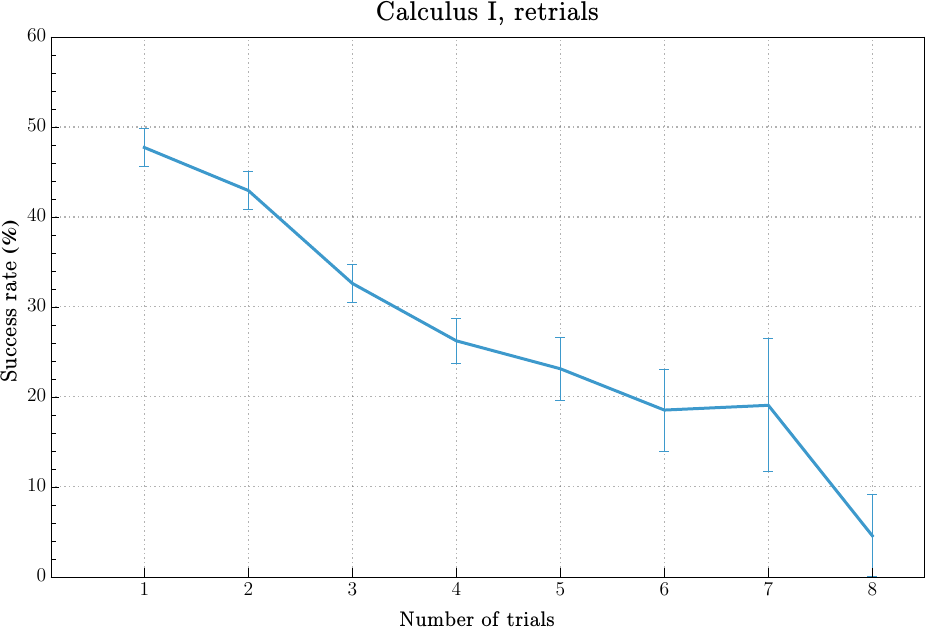}%
\caption{Success rate of students as they try to pass Calculus I consecutive times, between years 2015 and 2024. Each dot is the ratio $P_n \over E_n$, where $P_n$ the number of students passed after trying $n$ times and $E_n$ is the total number of students that tried at least $n$ times.\label{aprov_calc_retry}}
\end{figure}

Figure \ref{aprov_calc_retry} shows the probability of passing Calculus I as a function of the number of attempts. The result is counterintuitive: instead of increasing with repeated exposure to the course, the success probability actually decreases with the number of trials, suggesting that students who fail repeatedly tend to accumulate persistent difficulties rather than improve their chances of success.

This is an important result that must be taken into account when evaluating quota policies. If repeated attempts systematically reduce the probability of success, students entering with weaker academic preparation -- as is often the case in quota subgroups -- may face compounded difficulties over time. Beyond the initial performance gap, the dynamics of repeated failures can also amplify evasion.   

\section{Conclusions}
\label{conclusion}

The analysis of more than a decade of data from Unesp reveals that the evaluation of quota policies requires careful methodological distinctions. A central aspect is the separation among the Humanities, the Physical and Mathematical sciences, and the Biological sciences. Aggregated indicators, dominated by areas with permissive pass rates such as Pedagogy, tend to mask differences that become evident when one considers more selective programs such as Physics and intermediate cases such as Biology. It would be important also for researchers in the Humanities to propose appropriate metrics for the evaluation of student performance, as approval in early courses or the overall graduation rates are making all students appear equally good.

The level of competition in the university admission is also a relevant factor. In programs with very low applicant-to-seat ratio, as observed in Physics, the presence or absence of quotas has little practical effect: most applicants are admitted regardless of the track, and graduation rates among groups are statistically indistinguishable. Biology, on the other hand, with a higher competition, exhibits systematic differences: SU and EP students consistently reach higher completion rates than PPI students. In that direction, it would be very interesting to have more studies that look into the correlation between the cut-off grade for admission in the programs, the longer term performance of students and the impacts of quota systems.

The role of statistics is decisive. With limited sample sizes, standard deviations are large, and results are not statistically significant once Poisson uncertainties and systematic database errors are included. Only with sufficiently large datasets, as in the combined analysis of Calculus I across Unesp’s STEM programs, do systematic trends become visible. In this case, the hierarchy of approval rates -- highest for SU, intermediate for EP, and lowest for PPI -- emerges clearly, confirming that differences are not artifacts of limited statistics. The analysis of Calculus I also reinforces the hypothesis that the higher competition amplifies the disparity between admission tracks, since the inclusion of STEM programs, more competitive than Physics, increased not only the overall performance but also the gap among groups.

Equally revealing is the behavior of students across repeated attempts in gateway courses. Instead of improving with time, the probability of passing decreases as the number of attempts increases, indicating that repeated failure tends to reinforce difficulties rather than promote mastery. It would be highly advisable for Unesp to look closely into this, and studies evaluating the short and long term psychological effects of repeated failing are necessary. This longer term correlation between admission and graduation is also present in the Biology program plot of Figure \ref{evol_bio}. If there was a ``catch up'' effect happening, with less prepared students at admission getting better with time and eventually performing as well as the rest, one would see a delay, as PPI students would take longer to graduate, but the same plateau would be reached in the long term, which is not the case. This suggests that gaps present at admission can be hard to overcome within higher education, particularly for students who start with significant deficits in prior preparation. In this context, more competitive admission processes, with higher applicant-to-seat ratios and rigorous selection, appear to be closely associated with better graduation outcomes. 

It is also important to emphasize that academic performance is not univocally determined by public schooling or racial categories within the Black, Brown, and Indigenous even though the initial selection into this category is based on phenotypical traits. Because candidates with higher entrance-exam scores within PPI or EP are reallocated to groups with higher cutoff scores, the effective correlation that emerges is primarily linked to student performance in the admission exam. In this sense, racial self-identification is a necessary condition for placement into a given quota subgroup, but it is not sufficient. The best-performing students in those social groups will be in the SU category, thus depleting both the PPI and EP of high-scoring students. Therefore, the results presented here should not be used to evaluate the academic performance of racial categories or of students from public schools.

The effectiveness of quota policies cannot be judged solely by the demographic composition of incoming classes. It must also be evaluated through academic trajectories, which depend strongly on field-specific rigor. Without such careful analysis, there is a risk of drawing overly optimistic conclusions about the success of policies that broaden access but may ultimately fail to generate a positive social impact, both in terms of graduating well-prepared professionals and safeguarding the reputation of universities.

\bibliography{quota}
\end{document}